\documentclass[preprint]{aastex}
\eqsecnum
\tightenlines
 \received{28 August 2000}
 \revised{08 January 2001}
 \accepted{10 January 2001}
 \journalid{VOL}{JOURNAL DATE}
 \articleid{START PAGE}{END PAGE}
 \paperid{MANUSCRIPT ID}
 \cpright{TYPE}{YEAR}
 \ccc{CODE}

\newcommand \eg {{\it e.g., }}

\newcommand \ie{{\it i.e.,}}

\newcommand \tAO {${AO}$}
\newcommand \tPSF 	{${PSF}$}		

\shortauthors{Sivaramakrishnan {\it et al.}}
\shorttitle{Adaptive Optics Coronagraphy}

\begin{document}

\title { Ground-Based Coronagraphy with\\ High Order Adaptive Optics }

\author{Anand Sivaramakrishnan}
\affil{Space Telescope Science Institute\\
3700 San Martin Drive, Baltimore, MD 21218}
\email{anands@stsci.edu}

\author{Christopher D. Koresko\altaffilmark{1}}
\affil{ Division of Geology and Planetary Sciences\\ California Institute of Technology, Pasadena CA 91125}
\altaffiltext{1}{Current address: Jet Propulsion Laboratory, 4800 Oak Grove Drive, Pasadena CA 91109}

\author{Russell B. Makidon}
\affil{Space Telescope Science Institute\\
3700 San Martin Drive, Baltimore, MD 21218}

\author{Thomas Berkefeld\altaffilmark{2}}
\affil{ Max-Planck Institut fur Astronomie\\K\"{o}nigstuhl 17, D-69117 Heidelberg, Germany}
\altaffiltext{2}{Current address: Kiepenheuer - Institut fur Sonnenphysik, Sch\"{o}neckstrasse 6, D - 79104 Freiburg, Germany}

\and
\author{Marc J. Kuchner\altaffilmark{3}}
\affil{ Palomar Observatory\\ California Institute of Technology, Pasadena CA 91125}
\altaffiltext{3}{Current address: Harvard-Smithsonian Center for Astrophysics, Mail Stop 20 60 Garden Street,  Cambridge, MA 02138}

\begin{abstract}

We summarize the theory of coronagraphic optics, and identify a
dimensionless fine-tuning parameter, $\cal F$, which we use to describe
the Lyot stop size in the natural units of the coronagraphic optical train
and the observing wavelength.
We then present simulations of coronagraphs matched to adaptive
optics (\tAO)
systems on the Calypso 1.2m, Palomar Hale 5m and Gemini 8m telescopes under
various atmospheric conditions, and identify useful parameter ranges for
\tAO\ coronagraphy on these telescopes.
Our simulations employ a tapered, high-pass filter in spatial frequency
space to mimic the action of adaptive wavefront correction.
We test the validity of this representation of \tAO\ correction by comparing
our simulations with recent $K$-band data from the 241-channel Palomar Hale 
\tAO\  system and its dedicated PHARO science camera in coronagraphic
mode.
Our choice of monochromatic modeling enables us to distinguish between
underlying halo suppression and bright Airy ring suppression in the final
coronagraphic images.
For a given telescope--\tAO\ system combination, we find that \tAO\ systems
delivering images with Strehl ratios below a threshold value are not
well-suited to diffraction-limited coronagraphs.
When Strehl ratios are above this threshold, an optimized
coronagraph with occulting image plane stops as small as $4\lambda/D$
create a region around the \tAO\ target where dynamic range is significantly
enhanced.

\end{abstract}

	    

\section{Introduction}\label{introduction}

Many ground-based telescopes use adaptive optics
(\tAO) to produce diffraction-limited images at near-infrared
and visible wavelengths.  The point spread function (\tPSF)
of a telescope using \tAO\ typically consists of a bright,
diffraction-limited core with several Airy rings superimposed on a
wide scattered light halo containing several percent of the
total flux.  The improvement in image quality that \tAO\ provides
makes it possible to study the region within a few times the
diffraction width of the image of a bright star, with dynamic range
limited by the presence of the halo and bright Airy rings rather than
by atmospheric seeing.  

Present-day astronomical \tAO\ systems routinely deliver Strehl ratios 
of 30--70\%  (the Strehl ratio is the ratio of the peak intensity of the image
to the peak intensity of the image if the wavefront were free of all
aberration).  In this paper we look at what is possible when
optimized coronagraphs are used  with \tAO\ systems delivering 
50--95\% Strehl ratios.
We believe that this should be the next scientifically-driven step for 
traditional \tAO\ systems, since the instruments we model here will
open up a new
range of ground-based astronomical investigation, and provide a stepping
stone to even higher dynamic range ground-based astronomy
that is outside the reach of  multi-conjugate
\tAO\ systems used to correct wide fields of view \citep{berke01a}.

A coronagraph used in conjunction with an \tAO\ system can 
improve the sensitivity of an imaging system to faint structure surrounding 
a bright source.  This device blocks the core of the image of an
on-axis point source and suppresses the bright diffraction rings
and halo, removing light which would otherwise 
reduce the dynamic range of the imaging.
This enables faint off-axis structure to be observed.

Initial observational investigation of the dynamic
range achievable with a coronagraph on the Palomar Hale 5m \tAO\  system has 
demonstrated the importance of understanding the interaction between the \tAO\  system
and coronagraphic instrument parameters \citep{bro2000spie}.
The choice of pupil plane and image plane stops sizes
can only be made by understanding image formation by the 
\tAO\  system and the interaction between the \tAO\ system and 
the coronagraph.
A traditional coronagraph reduces off-axis light from an on-axis source
with two optical stops: an occulting stop in the image plane, and a matched
Lyot stop in the next pupil plane citep{lyot}.
An alternative solution to this problem is a nulling coronagraph
(\eg\ \citet{guyon99}),
although this novel approach still needs technological development
in order to be applied to broad-band imaging.  The discovery space of a nulling
coronagraph is also somewhat complementary to the optimized \tAO\ coronagraphs
we describe here.

With an \tAO\ system in place, the image plane stop can be made 
very small: only slightly larger than the diffraction spot
itself.  However, a pupil plane stop
matched to a reduced image plane stop must have a small clear diameter to
significantly reduce the off-axis throughput;  there is 
a trade-off between throughput and scattered light suppression.
In some situations using a coronagraph does not 
improve the final dynamic range of an instrument --- it 
simply extends the exposure time required for a given amount of
detected signal.

To address this issue, we simulated image formation in a coronagraph
mounted on a telescope with an \tAO\  system, and investigated the effects 
of atmospheric turbulence under a range of seeing conditions, telescope
sizes, and \tAO\ system performance levels.  
We introduce the principles of
coronagraphy with a simple one-dimensional model, and discuss the
results of the simulations in the context of 
current and next-generation instruments. 

Our model is an extension of that of \citet{tn94}, who modeled
low order ground-based \tAO\  coronagraphs
to determine the detectability of Jupiter-mass companions
around main sequence stars.  Nakajima simulated the \tAO\ wavefront
correction by zeroing out the lowest order Zernike coefficients 
in the expression 
used to generate realizations
of Kolmogorov-spectrum fluctuations in the atmosphere's refractive index.
In Nakajima's work, the Lyot stop is ``oversized'' by a fixed amount 
(10\%), and occulting stop sizes are chosen to be 1, 5, 10 and
15 times the telescope resolution.  We extend Nakajima's approach
in four ways.  First, we match the Lyot stop oversizing to the
size of the image plane stop to optimize coronagraphic performance.
Second, we use a graded high-pass filter to mimic \tAO\ correction.
Third, we investigate higher order adaptive correction of the
incoming wavefront to model 
\tAO\ systems with a few thousand actuators.
We concentrate on the smallest stop
sizes allowed by Fourier optics that do not lead to an unacceptable
reduction in throughput.
In earlier work \citep{rbm2000spie} we investigated occulting
stop sizes between $3 \lambda/D$  and $6 \lambda/D$ radians.   
Here we present results for a single occulting image plane stop 
with a diameter of $4 \lambda/D$ radians.
Present-day computer memory and processor speeds, as well as the development of
a rapid Markov algorithm to create the Kolmogorov-spectrum phase screens
used to simulate atmospheric effects \citep{glindem93, berke01b} 
enabled us to generate 1000 realizations
of each instrument configuration.  We therefore simulate long exposure
\tAO\ images which can be used to estimate detection limits,
since speckle noise is often the limiting factor in faint companion detection
\citep{racine1, racine2}.

\section{ Theory } \label{theory}

To set the stage for discussion of a more realistic
instrument, we illustrate the Fourier optics
of a one-dimensional coronagraph.
Other more formal expositions of coronagraphic imaging can be
found in the literature \citep{noll, vaughan, malbet}.  
Our analysis assumes that the Fraunhofer approximation
applies, \ie~that the transverse electric field in the image plane
is the Fourier transform of the phasor
of the wavefront phase in the pupil plane
	(if $\phi(x,y)$ is the phase, then $e^{i\phi(x,y)}$ is the 
	corresponding phasor).
We make use of the standard Fourier analysis results which can be found in
\citet{bracewell}.

For this analysis we consider only monochromatic imaging, but note that in 
typical broad-band imaging, the final image can be described
by the sum or integral of several monochromatic images, weighted by
the instrumental transmission function.
Wavelength variation across the band will act in
such a way as to smear image features radially by the same
factor as the fractional bandwidth, since the wavelength
enters into diffraction-limited image formation only in the
combination $(\lambda/D)$.  As a result, bright Airy rings
will get wider, but coronagraphic suppression of such 
rings will persist.  By treating the monochromatic case
we can distinguish clearly between halo suppression and Airy ring suppression.
Secondary support spiders and scintillation (field strength variation)
are not modeled here.


\subsection{The one-dimensional coronagraph without an atmosphere}

In the absence of atmospheric degradation, a monochromatic
on-axis source at infinity produces a transverse electric field at the 
telescope pupil
    \begin{equation} \label{in_wave}
	E\,=\,E_{0} \, Re(e^{i(kz-\omega t)}),
    \end{equation}
    where $k = 2\pi/\lambda$, the $z$ axis is the optical axis, 
    $\omega$ the angular frequency of the wave, $t$ the time, and $Re()$
    denotes the real part of a complex number.

We follow the passage of the incident wave's  field through
a one-dimensional coronagraph.  Figure 1 shows a diagram of
the optical path.

\begin{figure}
\figurenum{1}
\epsscale{0.95}
\plotone{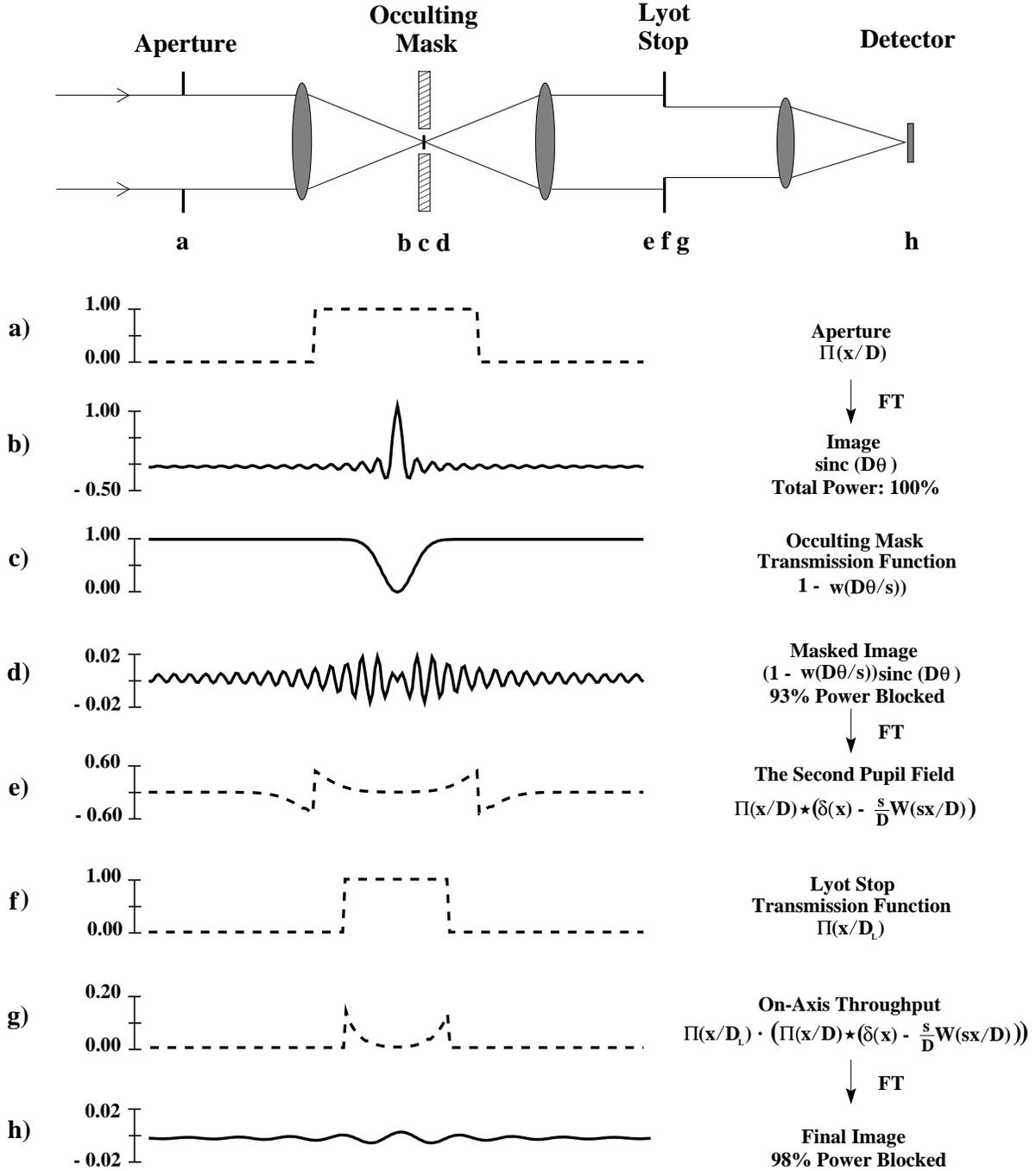}
\figcaption[fig1.eps]
{One-dimensional coronagraph summary, with locations and electric field
or stop profiles of:
    (a) primary pupil for on-axis source; 
    (b) image before image plane stop; 
    (c) image plane stop;
    (d) image after image plane stop;
    (e) pupil before Lyot stop;
    (f) Lyot stop;
    (g) pupil after Lyot stop;
    and (h) final on-axis image.
In this example, 98\% of the incident power is blocked by the coronagraph.
\label{fig1}}
\end{figure}

\placefigure{fig1}

We label eight key locations along this path with letters a--h.
Eight plots in the figure show 
the electric field 
due to an on-axis source at these key locations, and
the transmission functions of the optical stops
that affect the incident wave as it passes through the coronagraph.

First, the incoming wave passes through the telescope aperture 
(Figure 1a). We represent this interaction by multiplying
the field by the aperture stop function, so that
in the pupil plane
    \begin{equation} \label{aptophat}
	\begin{array}{ll}
	    E_{a} \,=\, E\,  \Pi(x/D_{\lambda}), & {\rm where }
	\end{array}
	\begin{array}{ll} \Pi(x)\,=\,1 & {\rm for\ } |x| < 1/2, \\
	                  \Pi(x)\,=\,0   & {\rm elsewhere}.
	\end{array}
    \end{equation}
Here $D_{\lambda}\,=\,D/{\lambda}$ is the number of wavelengths across
the telescope aperture 
	(which is also the inverse of the angular resolution of the telescope 
	as measured in radians).
	We denote a pupil plane coordinate by $x$, and an image plane
	coordinate by $\theta$.
The telescope optics then form the wave into an image
(Figure 1b).  The electric field in the image plane is the
Fourier transform of the aperture field $E_{a}$:
    \begin{equation}
        E_{b} \,\propto\, {\rm sinc}(D_{\lambda} \theta),
    \end{equation}
where $\theta$ is the field angle in radians in the first image
plane.  We omit the constants of proportionality 
for simplicity.  

In a conventional imaging camera this image field would fall on
a detector here.
However, in a coronagraph the star is occulted by a
field stop in this image plane.
We describe the stop in terms of a shape function 
$w(D_{\lambda}\theta/s)$, which is unity where the stop
is opaque and zero where the stop is absent.
If $w(\theta)$ has a width of order unity, the stop size
will be of the order of $s$ resolution elements.
The transmission function in the image plane is therefore 
$1 - w(D_{\lambda}\theta/s)$ (Figure 1c).
To illustrate the present discussion, we take
$w(\theta)~=~ {\rm exp}(-\theta^2/2)$.
The field in the first imaging plane after the occulting stop 
(Figure 1d) can therefore be written as

    \begin{equation} \label{postimstopfield}
 	E_{d} \,\propto\, {\rm sinc}(D_{\lambda}\theta)
	    (1 - w(D_{\lambda}\theta /s)).
    \end{equation}

This occulted image is relayed to a detector through a second pupil plane.
The electric field at this second pupil is the Fourier transform 
of the occulted image field (see Figure 1e):

    \begin{equation} \label{lyotstopfield}
 	E_{e} \,\propto\, 
		\Pi(x/D_{\lambda}) * (\delta(x) - {s \over D_\lambda}\  
			W(s\,x/D_{\lambda}))
    \end{equation}

Here $W$ is the Fourier transform of the image stop function
$w$, $\delta(x)$ is the Dirac delta function, and $*$ denotes
convolution. 
$W$ has width of order unity, although it will not have
bounded support
for occulting stop shape functions of finite extent, such as 
the hard-edged stop $w(\theta) = \Pi(\theta)$
	(the support of a function is the set of points at which
	the function is non-zero).
The geometrical significance of (\ref{lyotstopfield}) becomes clear if we
rewrite it as

    \begin{equation} \label{lyotstopfield2}
 	E_{e} \,\propto\, \Pi(x/D_{\lambda}) - 
	  {s \over D_\lambda}\ \Pi(x/D_{\lambda}) * W(s\,x).
    \end{equation}

If the image stop is completely opaque at its center,  $w(0) = 1$.
This means that its transform, $(s/D_\lambda)\ W(sx/D_\lambda)$,
has unit area, regardless of any re-scaling of
the argument of $w$.
This makes for cancellation of the field across most of the pupil 
when $s >> 1$.
In Figure 2 we show how the equation \ref{lyotstopfield2} is
constructed graphically, using a Gaussian image stop whose width 
is $5\lambda/D$ ({\it i.e.}\ ~$s = 5$).
This shows why the Lyot stop must mask out a border
of order $D/s$ wide around the pupil boundary to produce
significant reduction in the throughput of unocculted light
from the on-axis source.
It is only at this stage that the coronagraph increases
the dynamic range of the final image.

\begin{figure}
\figurenum{2}
\epsscale{0.75}
\plotone{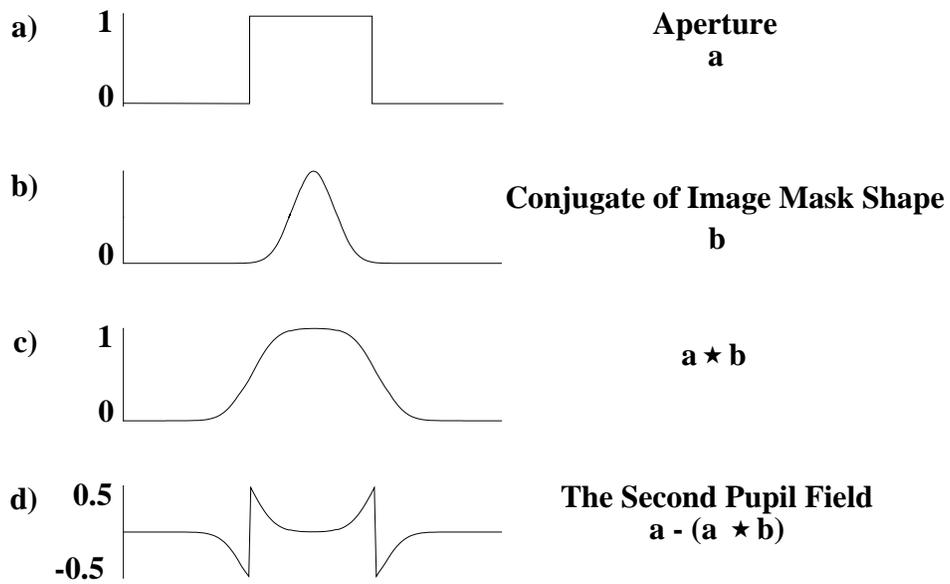}
\figcaption[fig2.eps]
{Graphical representation of the Lyot plane electric field calculation:
    (a) pupil function of width D;
    (b) Gaussian profile image plane stop with $5\lambda/D$ standard
    	deviation produces a Gaussian with standard deviation
        $D/5$ in the Lyot plane;
    (c) the convolution of the pupil function with the transform of the
        stop profile;
    (d) the final Lyot stop field showing bright edges and no energy in 
    the center.
\label{fig2}}
\end{figure}

\placefigure{fig2}

The unocculted light (Figure 1d) has a highly periodic distribution,
with periodicity $\sim \lambda/D$.  In the following pupil
plane (which is the transform space of the image plane), this energy
is concentrated near $\pm D/2$. The larger the occulting stop diameter,
the more $E_d$ looks like a pure sinusoid, and the more the unocculted energy
is localized in the neighborhood of the boundary of the following pupil.

In seeing-limited coronagraphs, the occulting stop is
typically many diffraction widths in size ($s \ge 10$). Consequently, the Lyot stop 
need only be undersized by a small fraction of the pupil diameter 
({\it e.g.}\  10\% or less),
resulting in minimal loss of throughput for unocculted, off-axis
sources.  In contrast, off-axis throughput in an optimized, diffraction-limited 
coronagraph with significant rejection of on-axis light must fall dramatically 
as the image plane stop shrinks to a few diffraction widths. 
This is because the spillover of unocculted on-axis light
occurs in a wide border around the pupil
boundary in the plane of the Lyot stop, so the Lyot stop must obscure 
a sizeable fraction of the re-imaged primary mirror to remove the
on-axis spillover.  Hence, off-axis throughput is reduced as well.

Since the scale of the Lyot stop oversizing is $D/s$, we fine-tune
the Lyot stop diameter so that it obscures a border ${\cal F} D / s$
around the perimeter of the primary.  The Lyot stop diameter
is therefore
    \begin{equation} \label{Lyotstoptuning}
        D_{Lyot} = D - 2 {\cal F} D/s.
    \end{equation}
If a secondary obstruction is present, then the Lyot stop must
block out a similar border around the inside edge of the
annular pupil.  This is why small secondary mirrors benefit
diffraction-limited coronagraphy.
In section \ref{optim} we describe how to choose an optimum value 
of $\cal F$ for a given telescope and \tAO\  system under given atmospheric
conditions.

Using the theory outlined above, 
in the the case of an unobstructed primary aperture with an image plane stop 
of $5 \lambda/D$ (\ie\  $s = 5$),
approximately  $16/25$ of the aperture should be obscured by 
a matched Lyot stop:
	when projected back onto the primary pupil,
	the Lyot stop is opaque outside a circle of diameter $ \sim 3D/5$.
Rejection of unwanted on-axis
light must be balanced by signal-to-noise considerations pertaining to the
off-axis source brightness.  This places a practical
lower limit on the angular size of the occulting spot
in the first image plane.

The above arguments hold for two-dimensional apertures as well.
The derivation is analogous to the one-dimensional case,
although the functions and transforms become two-dimensional
(\eg for a circular telescope aperture, 
the ${\rm sinc}$ function is replaced by the Airy function).
\citet{vaughan} describe the two-dimensional case, and 
\citet{malbet} treats the \tPSF's of
off-axis sources in such coronagraphs.


\subsection{ The coronagraph in the presence of an atmosphere }

In the absence of scintillation, we can model atmospheric effects on an
incident plane wave by multiplying the aperture illumination 
function $\Pi(x)$ by an atmospheric disturbance phasor $e^{\,i\phi(x)}$.
In our one-dimensional example, the resultant image field is
the Fourier transform of $\ e^{\,i\phi (x)} \Pi (x/D_{\lambda})$.\,
For telescope diameters larger than {$\sim0.5$}m at wavelengths shorter than 
$ \sim 2$ microns, the atmospheric 
phase function typically has significant power at length scales smaller
than the aperture size,  so a long exposure image under these
circumstances exhibits
the familiar unimodal seeing disk that is typical of
images from large ground-based telescopes.
In consequence, coronagraphs without \tAO\ systems on 
seeing-limited telescopes use occulting stops many diffraction widths wide,
with concomitant high throughput as explained earlier.
Stellar coronagraphy without adaptive wavefront correction,
or with only tip-tilt systems, is still useful to prevent detector saturation,
and has produced observations of considerable scientific  value
\citep{jhuaoc, gl229b_nature}.

Ideally, a coronagraph with an \tAO\  system would be just like
the coronagraph without an atmosphere.  The image of a point source
in the first image plane would be a pure Airy disk, and the size of the
Lyot stop could be chosen using simple Fourier theory and a diagram
like Figure 1.  However, an \tAO\  system cannot correct
atmospheric effects on spatial scales smaller than the inter-actuator
spacing in the pupil plane.  Power at high spatial frequencies that
goes uncorrected in the pupil plane transforms into noise on large angular
scales in the image plane.  The corrected image consists of a diffraction
limited core surrounded by an extended halo due to the uncorrected
aberrations, a remnant of the uncorrected seeing-disk.  The size and
shape of the halo reflect the number of actuators in the \tAO\  system and the
moment-to-moment characteristics of the turbulence in the atmosphere.

In the presence of this halo the choice of Lyot stop is not obvious.
Indeed, a given \tAO\  configuration may not be able to reduce the power in
the extended halo sufficiently to justify diffraction-limited
coronagraphy --- coronagraphy where the occulting mask is only a few
diffraction widths across.  In order to predict the size
of the uncorrected halo and to understand how it affects coronagraphy,
we modeled the effects of atmospheric turbulence on
the incoming wavefront and the correction of these aberrations by \tAO.
Our models allow us to find useful operating parameters for
diffraction-limited coronagraphs working in concert with \tAO\ 
systems for a range of telescope sizes, \tAO\  capabilities, and
observing conditions.  Sections \ref{optim} and \ref{thresh} discuss
these issues in some detail.


\section{Numerical Simulations}\label{simul}

Each of our numerical simulations is characterized by the telescope entrance
pupil size and geometry, the seeing, $D/r_o$, the number of actuators across
the primary, $N_{act}$, the size of the occulting image plane stop, $s$, 
the Lyot stop fine tuning factor, $\cal F$, and the linear size of the
array, $N_s$, sampling the incoming wavefront.
We generate 1000 independent realizations of Kolmogorov-spectrum
phase screens in $2N_s$ by $2N_s$ arrays, with the spatial sampling of these
arrays chosen to provide several samples across each Fried length $r_o$ 
\citep{fried},
while constraining the sampling to force $N_s$ to be a power of $2$.  
Our resultant spatial sampling ranges from $3.2$ to $21$ samples across $r_o$
(see Table 1).  

We Fourier transform the input phase arrays, and multiply them by a high-pass
filter to mimic the action of \tAO.  Then we reverse-transform
the filtered arrays to obtain the \tAO-corrected wavefront.  

The shape of the \tAO\  filter near cutoff depends on details of the
\tAO\  system.
Deformable mirror (DM) actuator influence functions which extend to neighboring
actuators' positions reduce the sharpness of the cutoff.  Noisy wavefront
sensing and intrinsic photon noise reduce the efficacy of high spatial
frequency wavefront correction.
The flow of the atmosphere past the telescope pupil and a non-isotropic
refractive index spatio-temporal distribution
also change the shape of the \tAO\  filter, as does imperfect DM calibration.

The cutoff frequency of the \tAO\  filter cannot be higher than the
spatial Nyquist frequency of the actuator spacing, 
	$k_{AO} = N_{act} / 2D$.
	At a given wavelength $\lambda$, this spacing corresponds to an angle
	\begin{equation} \label{thetaAO}
	\theta_{AO} = N_{act} \lambda / 2 D.
	\end{equation}
In earlier work \citep{rbm2000spie}, our \tAO\  high-pass filter was
the complement of a Hanning filter.
This has a continuous derivative everywhere, and a very smooth
approach to the  cutoff. 
As we show in Section 5, this proved to be too conservative,
as it  underestimates the amount of observed \tAO\ correction.
In these simulations we use a parabolic filter:
	\begin{equation} \label{parab}
	\begin{array}{ll}
	A(k) = (k / k_{AO}) ^2		& {\rm for }\  k < k_{AO}  \\
	A(k) = 1			& {\rm otherwise}. 
	\end{array}
	\end{equation}
We show that this matches observed results over the range where
comparison with data is valid.

To avoid edge effects introduced by the Fourier filtering,
only the central $N_s$ by $N_s$ section of the filtered array is retained.  This
is then multiplied by a binary mask representing the telescope entrance pupil.  
This mask is opaque (zero) outside the primary mirror edge, and inside the
secondary obstruction.  Secondary support spiders, mirror surface micro-roughness
and scattering are not considered in these simulations.

We embed the filtered and masked array in the center of
an $8N_s$ by $8N_s$ zero-filled array.
This results in image field sampling of $\lambda/8D$.  
which allows for effective comparison with 
data taken with $\sim \lambda/4D$ pixels.

We then create the complex phasor $e^{i\phi(x,y)}$ describing the electric
field corresponding to the phase $\phi(x,y)$, and Fourier transform the phasor
to obtain the first image field.  We average all 1000 realizations of the
intensity of this image, as well as the Strehl ratios derived from these
images as a measure the \tAO\  system performance.

This \tAO-corrected image field is then multiplied by another binary mask to
produce the effect of the image plane stop, and the product is
inverse-transformed to create the pupil field at the Lyot plane, where it
is multiplied by a third binary mask representing the Lyot stop.  The final image
is produced by Fourier transforming this field.  We average all realizations
of the final image intensity, along with the values of various throughput and
geometrical obstruction descriptors of the optical system.
These averaged images can be thought of as a single image with an exposure
time corresponding to a thousand times the speckle lifetime.  This is
equivalent to a $\sim 100$ second exposure at Palomar in the $K$-band.

\section{Optimizing Coronagraph Design}  \label{ocd}

In this section we present the results of numerical
investigations into when \tAO\  benefits coronagraphy, and 
how one optimizes the design parameters for a coronagraph used
in conjunction with an \tAO\ system.

\subsection{Optimizing the Lyot Stop}\label{optim}

In section \ref{theory} we defined the diameter of the Lyot stop as
	$D - 2{\cal F} D / s$,
where $\cal F$ is the Lyot stop fine tuning factor.
Here we show how the value of $\cal F$ is chosen to optimize
a coronagraph for a specific telescope under particular atmospheric
conditions.  We present the results of an exploration of a coronagraph
on the Gemini telescope with atmospheric turbulence described by 
$D/r_o = 30$.

\begin{figure}
\figurenum{3}
\epsscale{0.95}
\plotone{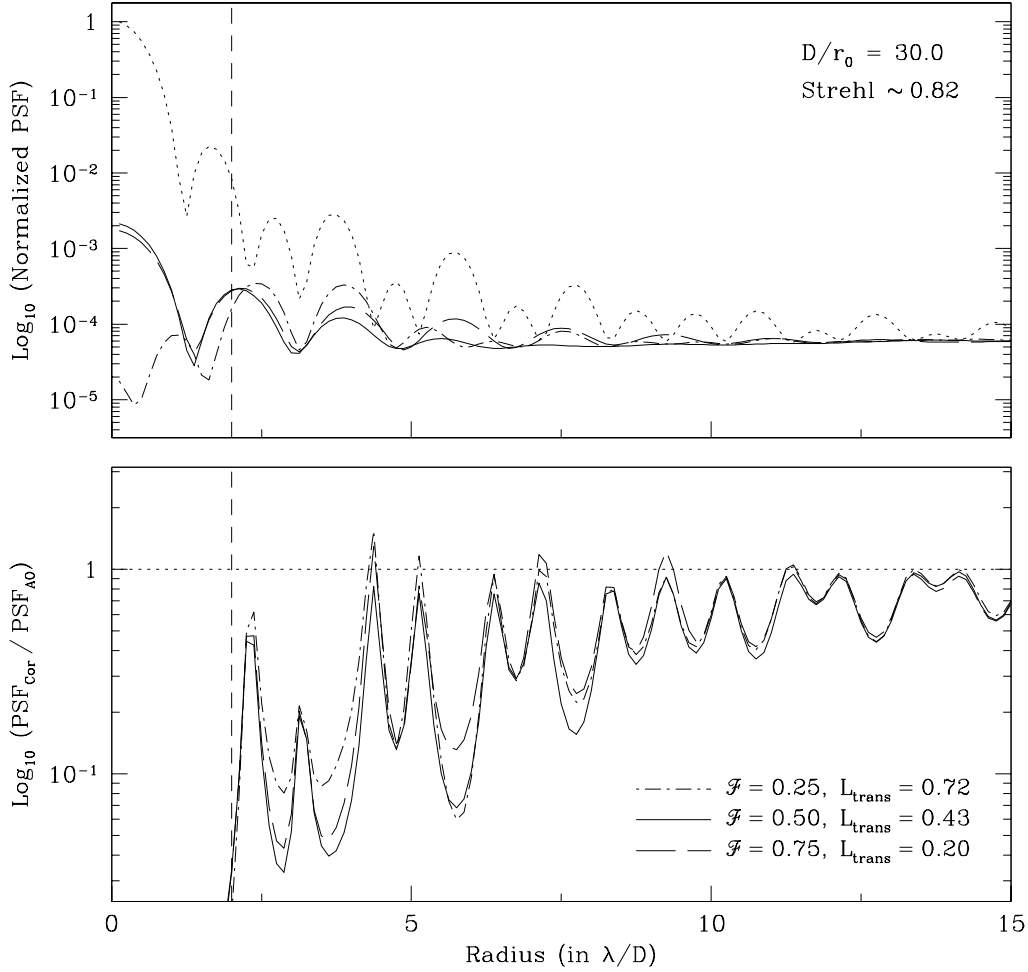}
\figcaption[fig3.eps]
{  Optimizing the Lyot stop diameter.
    Top panel: azimuthally averaged \tAO -only PSF
    (dotted line), and three coronagraphic PSF's with different
    Lyot stop diameters.
    The coronagraphic PSF's have been re-normalized to take Lyot stop losses
    into account (see text).  A $4 \lambda/D$ angular diameter occulting image
    plane stop, whose edge is indicated by a dashed vertical line in both
    panels,is used in these coronagraphic simulations.  Three values of the
    Lyot stop fine tuning parameter ${\cal F}$ describing the Lyot stop size
    are presented.
    Bottom panel: ratios of the azimuthally averaged coronagraphic PSF's to
    the \tAO -only PSF.	The most effective central image suppression is seen
    over a few diffraction rings outside the image plane stop for the
    intermediate value of ${\cal F}$.
\label{fig3}}
\end{figure}

\placefigure{fig3}

The uppermost curve in Figure 3a shows an
azimuthally averaged PSF that an \tAO\  system with 51 actuators across
the primary diameter (a total of 2042 actuators) delivers.  
A hundred independent atmospheric phase disturbance realizations have been
averaged into this PSF, which is normalized to unity at its center.
The Strehl ratio of the corresponding image is  82\%.

Below this we show azimuthally averaged PSF's of a coronagraph with 
a central stop of $4\lambda/D$ diameter, with three different values of
$\cal F$.  These curves have been re-normalized to take into account the
dimming of the image due to throughput losses that result from undersizing
the Lyot stop.  This is accomplished by dividing the simulated profile by
the Lyot stop clear fraction 
	(the ratio of the clear area of the Lyot stop, when projected back 
	to the primary, to the clear area of the entrance pupil).
If the \tAO -corrected PSF and a coronagraphic PSF
coincide at some particular value of the radial distance from the
central object, then, 
at that separation
there is no dynamic range gain to be had 
from using that coronagraph.  At such a location
the same companion signal-to-noise ratio will be
achieved when the same number of source photons have been detected.
In this case
the coronagraphic configuration simply increases the exposure 
time required to achieve the same signal-to-noise ratio, because it is 
equivalent to using a telescope with a smaller collecting area.

Figure 3b shows the ratios of the various renormalized, azimuthally 
averaged coronagraphic PSF's to the PSF with just the \tAO\  system and no
coronagraph.  Three values of the fine tuning parameter $\cal F$ are
presented: $0.25, 0.5$  and $0.75$.  The PSF for least aggressive Lyot
stop (${\cal F} = 0.25$) shows the least dynamic range benefit.
Bright re-imaged pupil edges are not sufficiently masked out, though the
Lyot stop is 72\% clear.  The most aggressive stop (${\cal F} = 0.75$)
masks out
much more of the bright pupil edges, but the Lyot stop is only 20\% clear.
This results in a drastic reduction in the brightness of off-axis sources.
The intermediate Lyot stop (${\cal F} = 0.5$) does a better job of
balancing the obscuring of bright pupil edges, and has a 
Lyot stop throughput of Lyot stop (43\%).
For a specific  instrument project this optimization will need to be done
on broad-band images, with finer resolution in the $\cal F$ parameter space.
We present this coarse optimization search as a model for further work
tailored to particular telescopes and science drivers.

\subsection{Threshold AO performance for coronagraphy}\label{thresh}

Here we consider the transition between seeing-limited and high 
Strehl imaging.  We select fixed seeing conditions ($D/r_o = 30$), 
and vary the actuator spacing of the \tAO\  system.
The resulting images have Strehl ratios between 30\% and 90\%
as the number of actuators across the primary diameter goes from 
26 to 71 (Figure 4).  
The validity of our \tAO\  correction algorithm is discussed
in section \ref{comparison}, we merely note here that it is based
on real data.  We normalize the PSFs in Figure 4 to be unity at the
center because we wish to present dynamic range gains rather than
absolute flux levels. 

\begin{figure}
\figurenum{4}
\epsscale{0.95}
\plotone{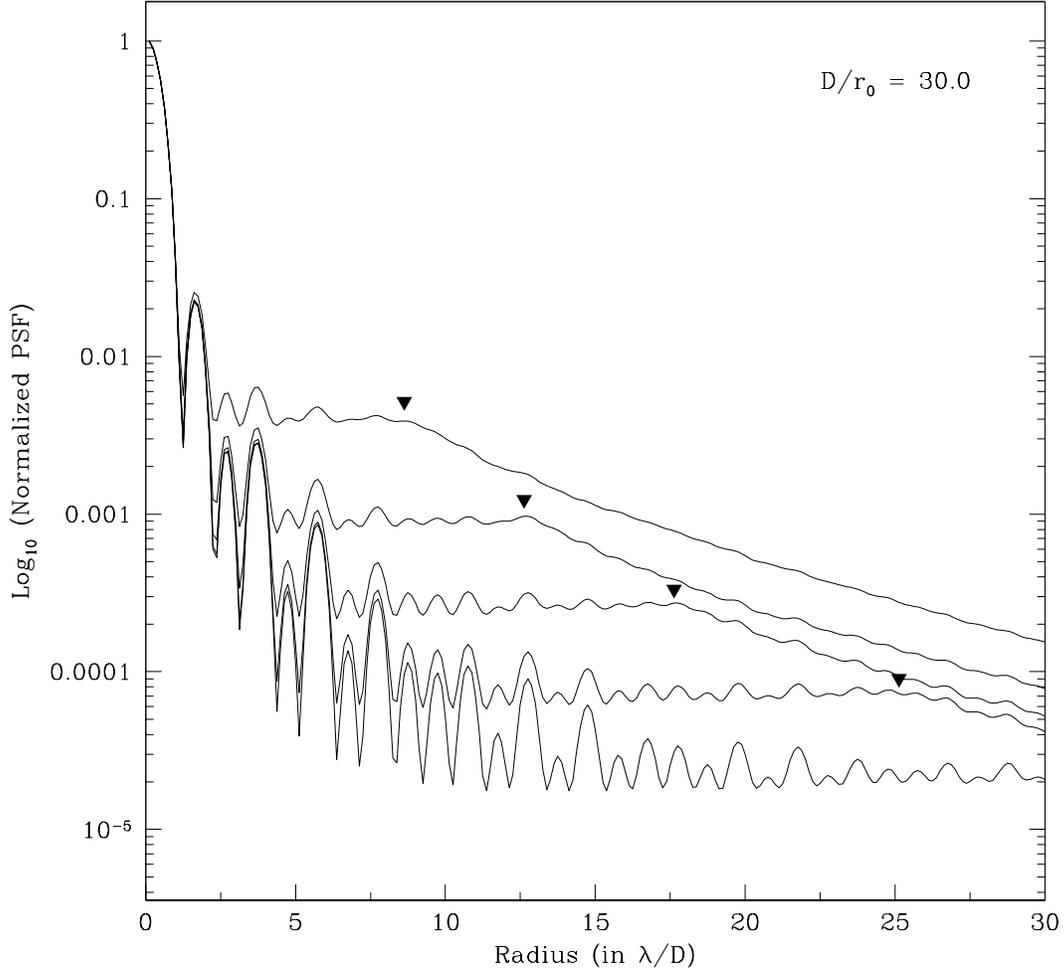}
\figcaption[fig4.eps]
{   The effect of increasing \tAO\  correction is seen here.
    The same 100 independent realizations of a Kolmogorov-spectrum
    phase screen representing atmospheric wavefront aberration 
    corrected by \tAO\  systems with 18, 26, 36, 51 and 71 actuators across
    a $D = 8.1$m primary mirror, with $D/r_o = 30$.
    The azimuthally averaged PSF's are normalized to unity at their center.
    The Strehl ratios are 0.30, 0.53, 0.69, 0.82 and 0.90 respectively.
    Note the widening plateaus of partial correction extending to the
    shoulder (solid triangle) at a distance of $\theta_{AO}$
    from the center of the image.
    Outside this plateau \tAO\ correction does not
    improve the image.
\label{fig4}}
\end{figure}

\placefigure{fig4}

The PSF is described by a diffraction-limited core, a flat \tAO--corrected
halo out to a shoulder at a radius $\theta_{AO}$ where the \tAO\  stops working.
After that the profile shows the familiar atmospheric wings.
The flat plateau of \tAO\  correction drops in intensity and grows
in radius as the number of actuators increases.  It is the uncorrected 
light in this area that is redistributed into Airy rings by the
\tAO\ system, and it is this light that is not removed by
a coronagraph.  The image with a Strehl ratio of 30\% has a
ten-fold decrease in the intensity of this scattered light
as compared with the 53\% Strehl image.  This demonstrates why there is a
threshold Strehl required for coronagraphic imaging targeted at a particular
angular distance from the central source.

Azimuthally averaged PSF's (dotted lines) for the same 100 realizations
of the atmospheric phase disturbances corrected by AO systems with a
range of actuator spacings
are shown in Figures 5a -- 5f.   Below these PSF's
we plot renormalized PSF's for a coronagraph with a $4\lambda/D$
occulting spot and a Lyot stop fine-tuning parameter of
${\cal F} = 0.5$.  The vertical scales on all these plots is the same,
and the PSF is normalized to unity at the origin.

The effect of increased \tAO\  correction on direct imaging is
manifest in the steadily decreasing intensity of the wings
relative to the central peak.

\begin{figure}
\figurenum{5}
\epsscale{0.95}
\plotone{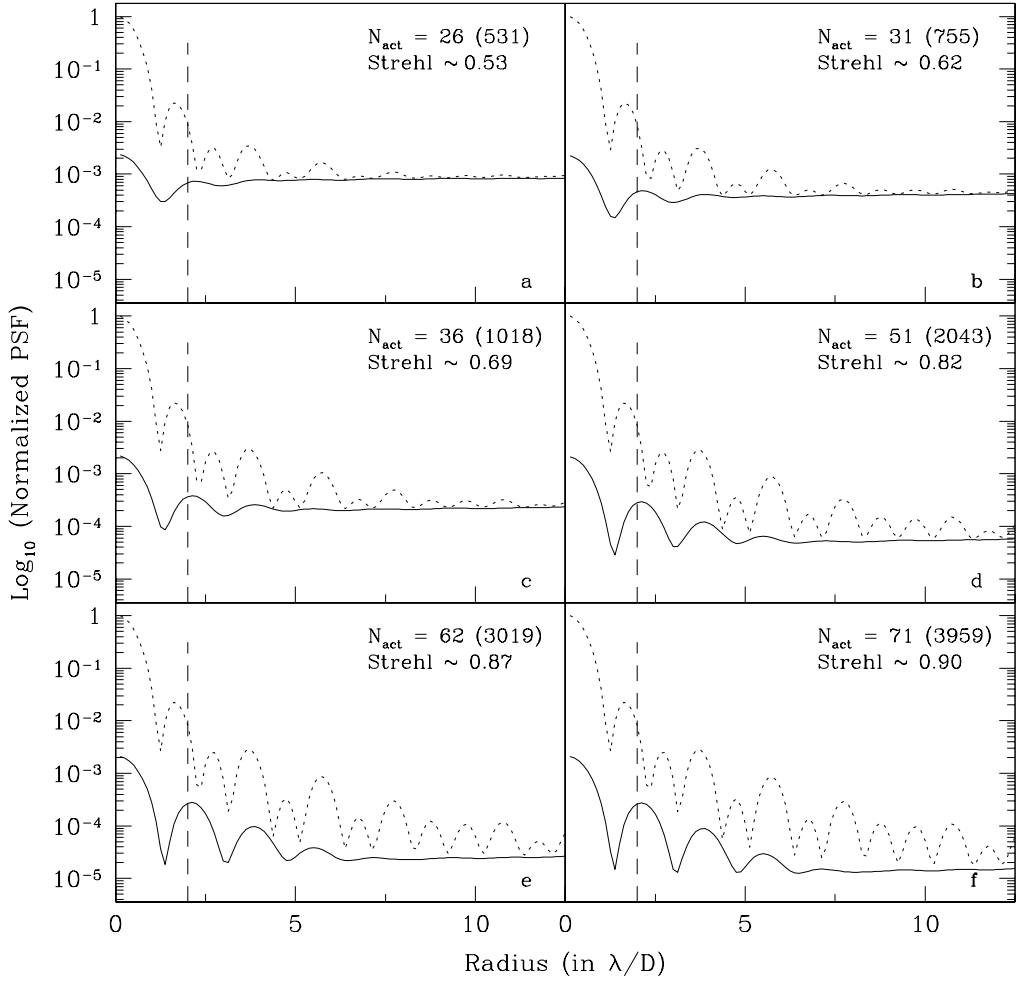}
\figcaption[fig5.eps]
{  The effect of varying degrees of \tAO\  correction on increased dynamic range
produced by a coronagraph with a $4 \lambda / D$  occulting stop  diameter.
The dotted lines are azimuthally averaged, co-added  PSF's for the same 100 realizations of
Kolmogorov-spectrum  atmospheric phase disturbances with $D/r_o = 30$. 
Solid lines are renormalized azimuthally averaged  co-added PSF's of the same
atmospheric phase realizations using a coronagraph with 
a Lyot stop fine-tuning parameter of ${\cal F} = 0.5$.
The vertical scales are identical, and the non-coronagraphic PSF's are normalized
to unity at the origin.  Renormalization is performed by dividing coronagraphic
PSF's by the fractional throughput of the corresponding Lyot stops (see text).
Panels a through f show the emergence of an annulus of image suppression
outside the occulting stop edge as the Strehl ratio increases from 53\% to 90\%.  
\label{fig5}}
\end{figure}

\placefigure{fig5}

At a Strehl ratio of 53\% (Figure 5a), the wings of both the
direct and  coronagraphic PSF's are about $10^{-3}$ of the
central intensity.  There is little suppression of the halo
of the image, though bright Airy rings are removed out to a
radial distance of $\sim 6\lambda/D$.  Scattered light from 
uncorrected atmospheric aberrations are not suppressed by
the coronagraph.  At a Strehl ratio of 62\% (Figure 5b),
AO correction produces visible Airy  rings further out,
since more light is guided into the Airy pattern and less
light is scattered semi-randomly into the halo.  Again,
residual scattered light is not suppressed coronagraphically,
and it still limits dynamic range.  In Figures 5c and 5d we see
further reduction in the scattered light halo, and a corresponding
increase of energy in the pure Airy pattern.  Coronagraphic
imaging suppresses the image to well below the darkest Airy rings,
a phenomenon we call halo suppression, since that is how it appears
to the observer, even though it is really suppression of a coherent
diffraction pattern emerging because of the reduction in
halo intensity.  

By the time the Strehl ratio has risen to 82\% (Figure 5d),
an annulus of greatly increased dynamic range opens up
between the occulting stop edge at a radius of $2\lambda/D$,
and the radius at which the remaining halo is comparable
in intensity to the Airy pattern, at a radius of $4\lambda/D$.
It is in such regimes that dynamic range increases dramatically.
Below such Strehl ratios a coronagraphic spot with a $4\lambda/D$
diameter would not produce much gain in dynamic range.

The improved dynamic range would continue indefinitely if \tAO\  guide
stars were bright enough.  However, for a given primary area,
as the number of \tAO\  channels increases, the number of photons
per channel
available to use for \tAO\  correction decreases.  We extrapolate from
Palomar \tAO\ system performance to the regimes shown in Figures 5e and
5f as goals that could be reached by the next generation of
\tAO\  systems on 8m telescopes at good astronomical sites.

\section{Comparison of Simulations with Palomar PHARO Observations} \label{comparison}
 
We compared our monochromatic models to a 180 second $K$-band exposure
obtained with the PHARO camera on the Palomar \tAO\ system,
using an image plane stop of $0 \farcs 96$ 
in diameter \citep{bro2000spie}.
We used the appropriate Lyot oversizing to match the
PHARO coronagraphic Lyot stop dimensions (${\cal F} = 1.07$, or a
clear aperture of 4m),
and match the \tAO\ system actuator spacing of 16 across the primary.
Details of the PHARO camera optics are described in \citet{hayward00}.

Figure 6 shows an overlay of the radial profiles of our
simulations with $D/r_o = 10$ and the parabolic \tAO\  high pass filter (dashed),
against the data (solid).  Both images were normalized to contain unity power
out to a radius of $4 \farcs 0$.

Our choice of $D/r_o$ results in a match with the wings of the
stellar image past 
$\theta_{AO} = 8\lambda/D$ (where the \tAO\  ceases to improve the image).
We did not need to fine-tune the Fried length $r_o$ to match
the observing conditions when the data were taken, as our initial guess at $r_o$
provided a sufficiently accurate estimate
of the azimuthally averaged stellar wing profile. 
Other effects
(non-Kolmogorov spectra, instrumental scatter, the effects of the secondary
support spiders, waffle or other wavefront reconstructor errors, {\it etc.})
could also be present in the PHARO data.

The simulated image plane stop has a diameter of $10.6\lambda/D$
at 2.2 microns.  Therefore the region where our model is tested by the data
lies between $5.3\lambda/D$ ($0 \farcs 48$) and $\theta_{AO} = 8\lambda/D$
($0 \farcs 73$).
Our model differs from observations by a few percent over the entire range where
the comparison is valid.  Detailed work on matching simulations with this data
will be presented elsewhere \citep{asbroinprep}.

The comparison with data suggests that our representation of
\tAO\ correction  as a parabolic filter is accurate enough to be used
to predict system performance of the next generation of \tAO\ coronagraphs.

\begin{figure}
\figurenum{6}
\epsscale{0.75}
\plotone{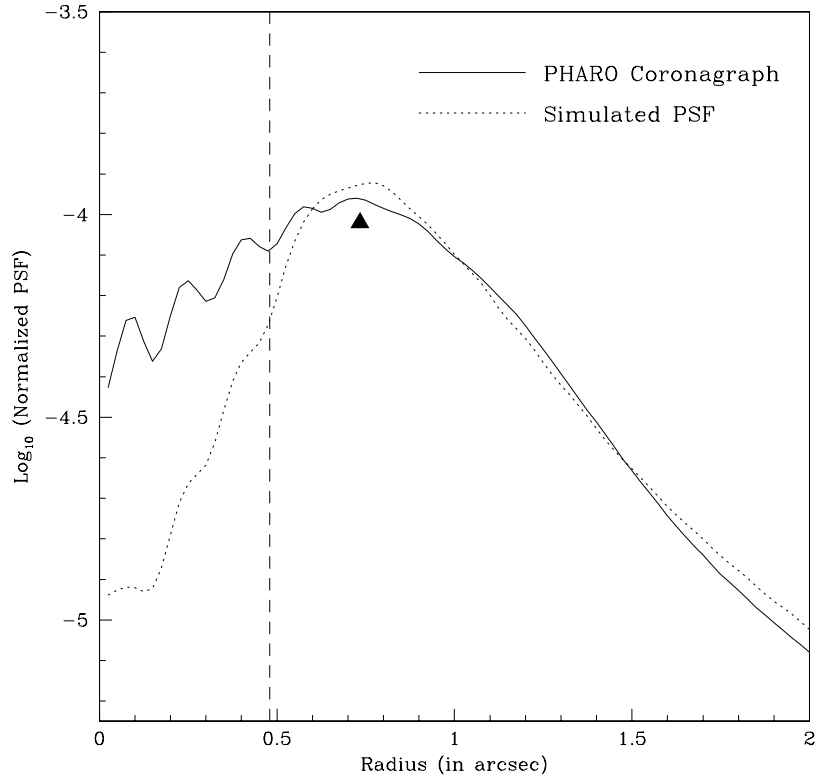}
\figcaption[fig6.eps]
{Azimuthally averaged image profiles of
    observed PHARO $K$-band data (solid) and
    and our simulations with $D/r_o = 10$,
    and a parabolic \tAO\  high pass filter (dashed).
    Both images were normalized to contain unity power
    out to $4 \farcs 0$.
\label{fig6}}
\end{figure}

\placefigure{fig6}

\section{DISCUSSION} \label{disc}

In order to span a range of telescope sizes, we simulated
the KPNO Calypso 1.2m, the Palomar Hale 5m, and the Gemini 8.1m.
We reproduce the secondary obscuration ratios of each telescope,
though we do not model secondary support spiders.

Atmospheric phase screen realizations using one or a few 
representative values of the Fried length relevant to the sites are
input to the optical train simulation after being smoothed by the \tAO\ 
filter. The effect of wind on the delivered Strehl ratio have not
been incorporated.

We match the \tAO\ system actuator density on the primary apertures to those
presented in previous work \citep{rbm2000spie}. A summary of the relevant
input parameters (primary and secondary mirror diameters, number of
actuators, and input $D/r_{0}$ values) can be found in Table 1.  

\placetable{table1}

While we examined a variety of image plane occulting stop sizes, we only
present one stop size, $4 \lambda/D$, in this paper.  Our
choice of this stop diameter is based on previous work which suggests  $4\lambda/D$
is close to the smallest effective stop size for imaging stellar coronagraphs
with Strehl ratios between 70\% and 95\% \citep{rbm2000spie}.
Larger stops simply obscure more of the improved image without a significant
increase in the dynamic range of the final images.
We find that stops smaller than $4 \lambda/D$ require extreme oversizing of the
Lyot stop, dramatically reducing system throughput. We note that for larger values
of the image plane stop diameter, the Lyot throughput penalty is
reduced.  In consequence, the optimal value of $\cal F$ is expected to grow
with increasing values of $s$ when $s \lesssim 10$.

Two values of $\cal F$, $0.5$ and $0.25$, are examined here. These yield
Lyot stop outer diameters of $0.75\,D$ and $0.875\,D$ respectively for 
our image plane stop. 
The effect of the two Lyot stops on the final image are illustrated in
the \tPSF\ plots of Figures 7, 8 and 9.
In general, we find a coronagraph coupled to the more aggressive Lyot stop
(${\cal F} = 0.5$) is significantly better at suppressing both Airy ring and
halo contributions to azimuthally averaged \tPSF s than the higher
throughput one (${\cal F} = 0.25$).

The coronagraphic \tPSF's presented in Figures 7, 8, and 9
take into account the Lyot stop throughput losses, as they have
been renormalized by dividing by the ratio of the Lyot stop area to the
primary aperture area.
When the image profiles with and without the coronagraph coincide,
no gain in dynamic range is attained by using a coronagraph.

We indicate the stop edge in these figures with a vertical line.
We also mark where the \tAO\ stops working (at the angular distance $\theta_{AO}$)
with a triangular symbol if this occurs within the plot boundaries.
In our Gemini simulations \tAO\ this angle falls outside the borders 
of the \tPSF\ plots.

\subsection{Kitt Peak Calypso 1.2m telescope} \label{calypso}

The Kitt Peak Calypso 1.2m telescope was designed and constructed to
minimize the deleterious effects of ground-level turbulence and scattered
light on astronomical observations \citep{smith99}.
This telescope's instrumentation operates in optical bandpasses.
 
For a small-aperture telescope like Calypso, we found that Strehl ratios of
at least 60\% are necessary for useful \tAO\ coronagraphy.
The simulation we present has a 75\% Strehl ratio.
\citet{smith99} reports upper quartile $r_o$'s of 20cm or less in
the $V$ band at this site ---  we use a value of 24cm.

The Airy rings are well-suppressed out to many diffraction widths
when ${\cal F} = 0.5$ (Figure 7a).
The broad underlying halo is reduced by about a factor of three
a few resolution elements past the image plane stop with this Lyot stop.

\begin{figure}
\figurenum{7}
\epsscale{1.00}
\plotone{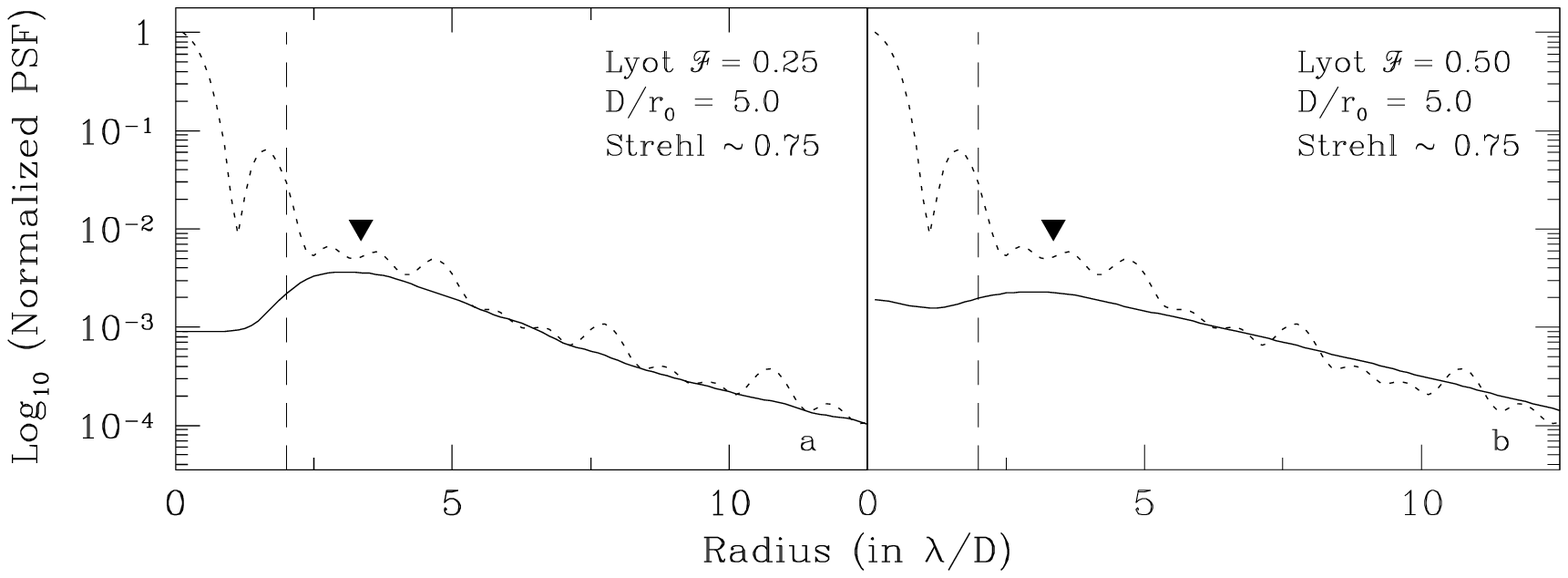}
\figcaption[fig7.eps]
{Azimuthally averaged \tPSF's for the 1.2m Calypso telescope,
    with 7 actuators across the pupil diameter ($\sim38$ actuators
    within the pupil)
    and an occulting image stop of $4 \lambda /D$.
    ($0\farcs 48$ at $\lambda = 0.7\micron$).
    The Lyot stop fine tuning parameter $\cal F$ is 0.25 on the left, and 0.5
    on the right (see text).
    The solid line shows the \tAO -corrected \tPSF, 
    the dashed line shows the image profile after the coronagraph.
\label{fig7}}
\end{figure}

\placefigure{fig7}

\subsection{Palomar Hale 5m telescope} \label{P200}

The Hale 5m has a functioning 241-channel \tAO\ system 
with a dedicated near-infrared science camera,  PHARO  \citep{hayward00}, 
which contains a variety of coronagraphic image plane and Lyot stops.

When $D/r_0 = 10$ we expect Strehl ratios of the order of $80\%$ 
(Figures 8a and 8b).  Such
conditions can be expected in the $K$-band on good nights.  At this
Strehl ratio the Airy ring structure is almost completely removed,
and halo is reduced by about a factor of two 
between the stop edge at $2\lambda/D$ and $\sim 5\lambda/D$ when 
${\cal F} = 0.5$ (Figure 8b).
The benefits of using a coronagraph when the \tAO\ delivers
a 50\% Strehl ratio are greatly diminished, as can be seen
by comparing figures 8b and 8d.  Thus the current 241-channel Palomar
\tAO\ system is suited to $K$-band coronagraphy under good conditions.
A system upgrade to \eg twice the number of actuators
will enable the coronagraph
to open up a larger search annulus in the $K$-band, and produce useful
dynamic range gains in the $H$-band.
To exploit the full potential of coronagraphic imaging on this system,
upgraded wavefront reconstruction
may be needed to delve into the cores of the images at angular separations
of the order of the stop radius of $2\lambda/D$ from the
target \tAO\ star.

\begin{figure}
\figurenum{8}
\epsscale{1.00}
\plotone{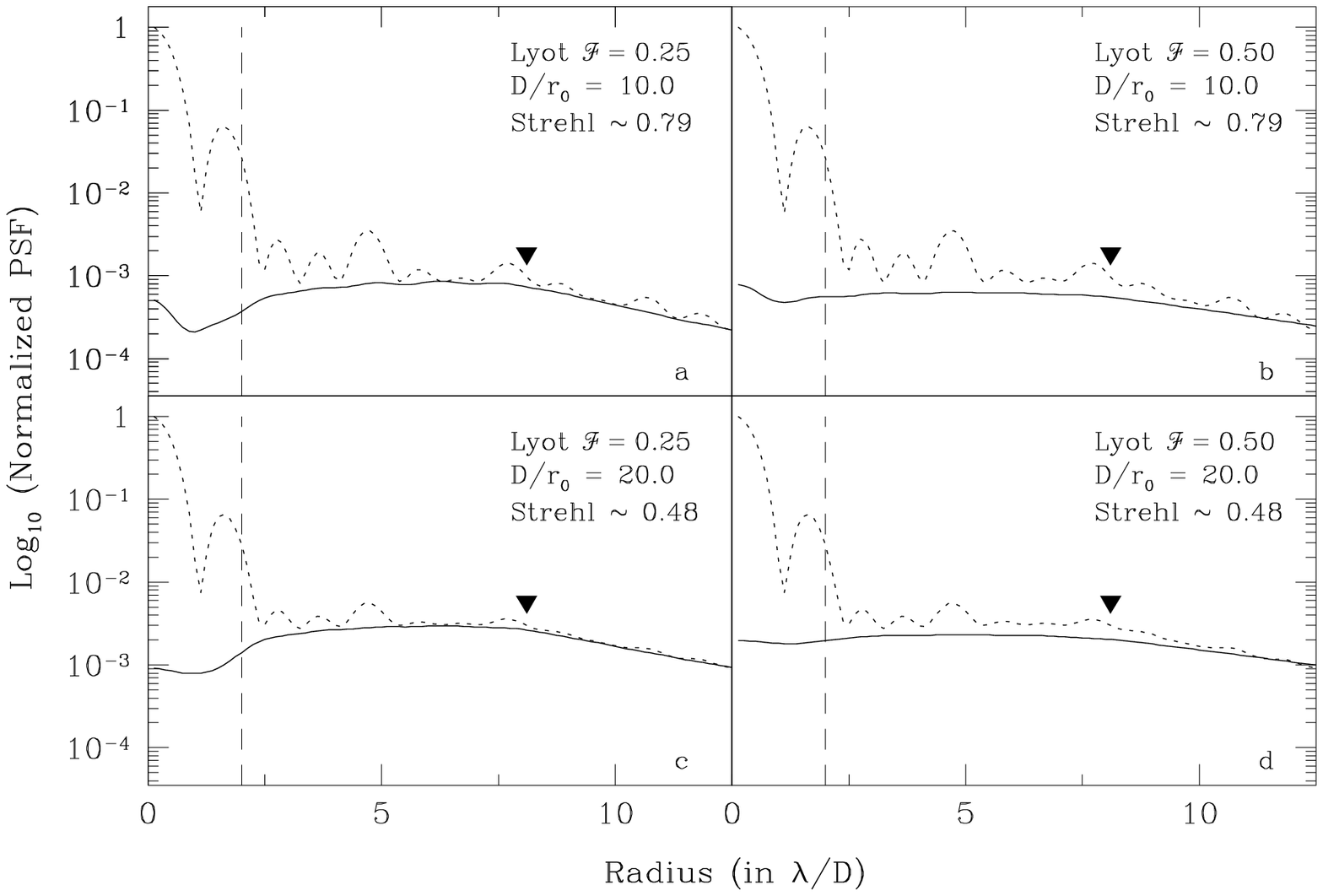}
\figcaption[fig8.eps]
{Azimuthally averaged \tPSF's for the 5m Palomar Hale telescope,
    with 16 actuators across the pupil diameter ($\sim216$ actuators
    within the pupil)
    and an occulting image stop of $4 \lambda /D$
    ($0\farcs 27$ at $\lambda = 2.2\micron$).
    The Lyot stop fine tuning parameter $\cal F$ is 0.25 on the left, and 0.5
    on the right (see text).
    The solid line shows the \tAO -corrected \tPSF, 
    the dashed line shows the image profile after the coronagraph.
\label{fig8}}
\end{figure}

\placefigure{fig8}

\subsection{Gemini 8.1m telescope} \label{gemini}

Gemini has a near-infrared coronagraph under construction, and 
will be equipped with a low order \tAO\ system, which enables
high order second stage \tAO\ \citep{angelnature} to be
incorporated into a dedicated \tAO\ coronagraph.
We used values of the Fried length which apply in the near-infrared.
Under these conditions, expected Strehl ratios of 80\% -- 95\% produce
extremely effective coronagraphic suppression.  In addition,
$\theta_{AO} = 25\lambda/D$,
so the annulus of high dynamic range imaging is relatively large
(Figure 9a through 9f).

With $D/r_o = 30$,  image halo reduction by a factor of three is observed
a few diffraction widths outside the stop edge, and Airy ring suppression
by another factor of three extends out to $\sim 5\lambda/D$ (Figure 9f).

At the other extreme, when  $D/r_o = 15$ (\eg in the $K$-band),
performance improves dramatically --- in Figure 9b
we see Airy ring suppression of a factor
of over 10 or more in the bright rings located at 3.5 and 5.5 $\lambda/D$,
and halo suppression of three out to $\sim 7\lambda/D$.

\begin{figure}
\figurenum{9}
\epsscale{1.00}
\plotone{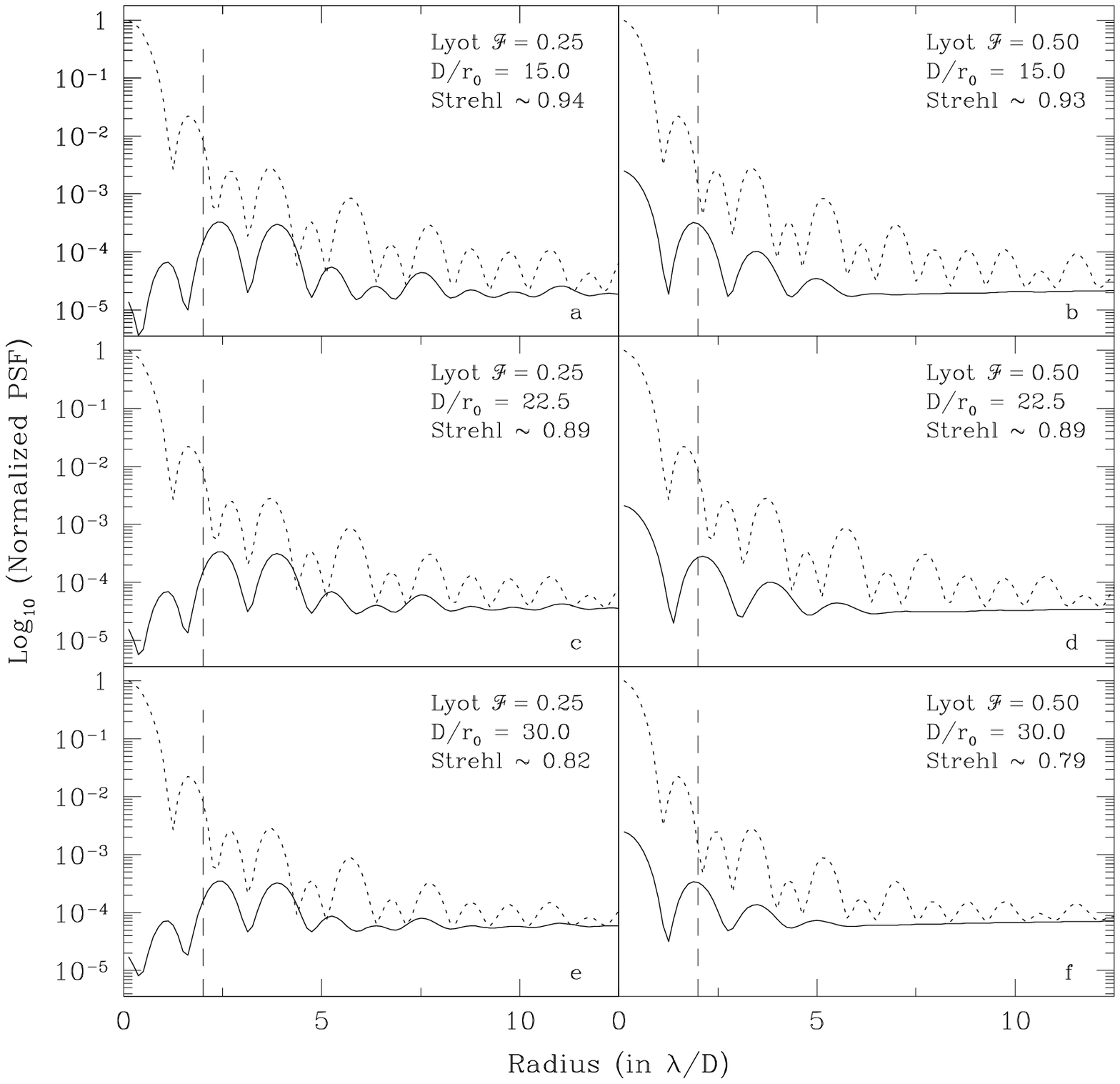}
\figcaption[fig9.eps]
{Azimuthally averaged \tPSF's for the 8.1m Gemini telescope,
    with 51 actuators across the pupil diameter ($\sim2042$ actuators
    within the pupil)
    and an occulting image stop of $4 \lambda /D$
    ($0\farcs 16$ at $\lambda = 1.6\micron$).
    The Lyot stop fine tuning parameter $\cal F$ is 0.25 on the left, and 0.5
    on the right (see text).
    The solid line shows the \tAO -corrected \tPSF, 
    the dashed line shows the image profile after the coronagraph.
\label{fig9}}
\end{figure}

\placefigure{fig9}

In Figures 10a -- 10d we show simulated
images for companions with brightness differences $\Delta m$ of
$+5$, $+7.5$ and $+10$ magnitudes relative to the on-axis target.  In the 
absence of noise, we see that a companion with $\Delta m = +10$
at a separation of $0 \farcs 16$ is observable  in 
the $H-$band under good seeing
conditions ($D/r_o = 15$) with a 2000 channel \tAO\ coronagraph
with a $0 \farcs 16$ diameter field stop (Figure 10d).  The exposure time 
corresponding to these simulated images is $\sim 1000$ speckle lifetimes.

\begin{figure}
\figurenum{10}
\epsscale{1.00}
\plotone{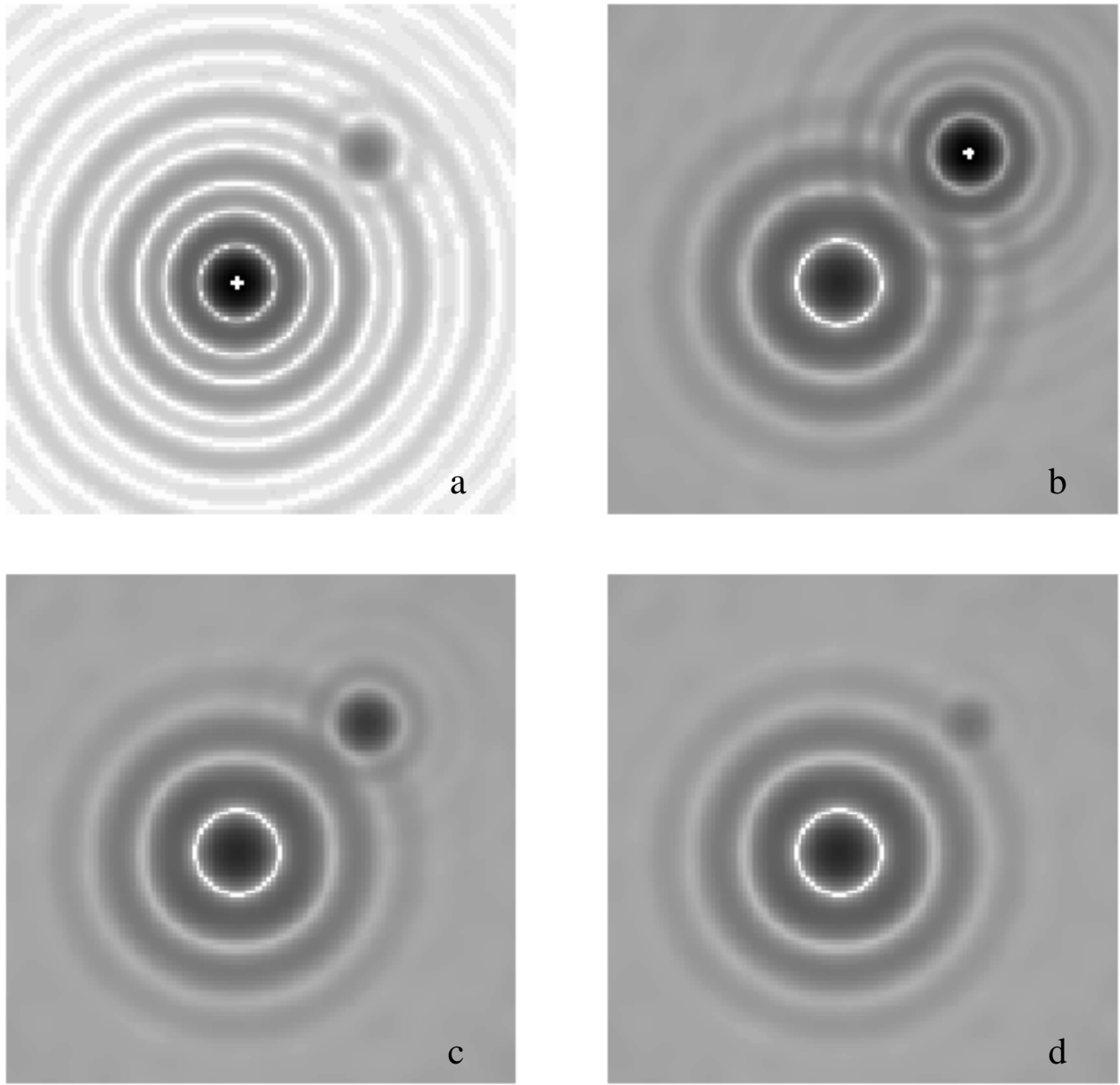}
\figcaption[fig10.eps]
{Realizations of a star with a faint companion at $4\lambda / D$
    separation with $D/r_o = 15$ for the Gemini 8.1m telescope assuming 
    2042 channels of \tAO.  In (a) we show the primary star and a companion 
    of $\Delta m\,=\,+5$ with no coronagraph in place. Figures (b), (c) and (d)
    exhibit companions of $\Delta m\,=\,+5$, $+7.5$ and $+10$ magnitudes
    respectively, with the primary occulted by an image plane stop of diameter 
    $4 \lambda /D$ ($0 \farcs 16$ in the $H$-band).
\label{fig10}}
\end{figure}

\placefigure{fig10}

\section{Conclusion} 

Diffraction-limited AO coronagraphy targets an entirely different
search space from seeing-limited coronagraphs.
Traditional coronagraphy under seeing-limited conditions
provides image suppression where the seeing halo drops below
the Airy pattern's wings \citep{bmac2000pc}.
Typically this is tens of diffraction widths away from the central bright object.
The advantage of diffraction-limited 
coronagraphic imaging is concentrated in an annulus
starting just outside the image stop, and ending where  the \tAO\ system
stops improving the wings of the \tAO\ target star's image.
The size of the outer edge of this annulus
is set by observing wavelength and the spatial Nyquist
frequency of the \tAO\ system actuators when projected on to the primary.
When small image plane stops are used the Lyot stop size must be
carefully chosen to find the best trade-off between throughput
and image suppression.  
Coronagraphic instruments might work extremely well in the near-infrared 
and longer optical bandpasses
on large telescopes with the next generation of deformable mirrors
using a few thousand actuators, provided wavefront correction is done well.

For 1-m class telescopes at $1\micron$,
we find that \tAO\  systems can deliver Strehl ratios $\ge 60\%$
without excessively compromising the brightness limits of available \tAO\ .
However, only modest contrast ratio gains are achieved with a coronagraph on
such telescopes. Coronagraphic suppression
is evident mostly in the bright Airy rings of the image.

The existing 241-actuator Palomar \tAO\  system is 
useful for $K$-band coronagraphy.  Stop sizes down to at least
$4\lambda/D$ are theoretically possible.
Improved wavefront reconstruction may be necessary  
for these small occulting stops.
With double the number of \tAO\ channels and reduced wavefront sensor
read noise, this telescope could deliver Strehl ratios of 90\%
at 2.2\micron.
$H$- and $K$-band coronagraphy would then become more scientifically
profitable because of the increased dynamic range in this search space.

On 8-m class telescopes, tremendous coronagraphic gains are to be had with 
2000 channel \tAO\  systems.  With occulting stop diameters of 
$\sim 0 \farcs 15$, three-fold halo suppression and ten-fold 
bright ring suppression can be expected a few diffraction widths from
the stop edge.
Such a system on the Gemini telescopes would be ideally suited to 
high dynamic range coronagraphy in the $H$-band,
with limiting Strehl ratios of $\sim$90\% possible
around an M0 dwarf \tAO\ target $\sim$20 parsecs away.

\acknowledgements

    We thank B. Brandl and S. R. Kulkarni for posing questions
    that led to this investigation, 
    and 
    P. E.  Hodge, T. Nakajima, B. A. Macintosh, and B. R. Oppenheimer 
    for insightful discussions.  
    B. R. Oppenheimer also provided us with pre-publication data.
    We are indebted to J. P. Lloyd for a critical reading of our simulation 
    code,
    and the anonymous referee for helpful suggestions. 
    We also thank R. J. Allen, J. R. Graham, E. P. Nelan, and R. L. White
    for encouraging the investigation.
    This work was funded in part by grants from the STScI Director's
    Discretionary Research Fund to AS and RBM,
    and the STScI Research Programs Office and visitors program for
    travel support for AS and CDK.

\clearpage

\begin{deluxetable}{llcccccccccc}
\tablecolumns{12}
\tablewidth{0pt}
\tablenum{1}
\footnotesize
\tablecaption{Properties of Telescope Simulations \label{table1}}

\tablehead{
\colhead{Telescope} & \colhead{} & \colhead {$D$} & \colhead{$D_{s}/D$} & \colhead{} & \colhead{$N_{act}$} & \colhead{$N_{chan}$} & \colhead{$N_{s}$} & \colhead{} & \colhead{$D/r_{0}$} & \colhead{} & \colhead{Strehl (\%)}
}
 
\startdata
 
Calypso &  &  1.2  &  0.36  &  &   7  &    38  &   64  &  &   5.0  &  &  75 \\
\cline{1-12}
Palomar &  &  5.08 &  0.33  &  &  16  &   201  &   64  &  &  10.0  &  &  79 \\
        &  &       &        &  &      &        &       &  &  20.0  &  &  48 \\
\cline{1-12}
Gemini  &  &  8.1  &  0.13  &  &  51  &  2042  &  128  &  &  15.0  &  &  94 \\
        &  &       &        &  &      &        &       &  &  22.5  &  &  89 \\
        &  &       &        &  &      &        &       &  &  30.0  &  &  80 \\
 
\tableline
\enddata
 
\tablecomments{For this table, 
    $D$ and $D_{s}$ are the primary and secondary mirror
    diameters (in meters).  
    $N_{act}$ is the number of actuators across primary mirror diameter,
    $N_{chan}$ the estimated number of \tAO\ channels within the pupil,
    and
    $N_{s}$ the number of wavefront phase samples across the primary
    mirror diameter.}
 
\end{deluxetable}

\clearpage

\end{document}